\documentclass[letterpaper]{article} 
\usepackage[submission]{aaai23}  
\usepackage{times}  
\usepackage{helvet}  
\usepackage{courier}  
\usepackage[hyphens]{url}  
\usepackage{graphicx} 
\usepackage{natbib}  
\usepackage{caption} 
\usepackage{algorithm}
\usepackage{algorithmic}

\usepackage{amsmath}
\usepackage{amsfonts}
\usepackage{graphicx}
\usepackage{subfigure}
\usepackage{multirow}

\usepackage{newfloat}
\usepackage{listings}
\DeclareCaptionStyle{ruled}{labelfont=normalfont,labelsep=colon,strut=off} 
\floatstyle{ruled}
\newfloat{listing}{tb}{lst}{}
\floatname{listing}{Listing}
\title{Consensus Learning for Cooperative Multi-Agent Reinforcement Learning}
\author{
    Zhiwei Xu, 
    Bin Zhang,
    Dapeng Li,
    Zeren Zhang,
    Guangchong Zhou,
    Hao Chen,
    Guoliang Fan
}
\affiliations{
    Institute of Automation, Chinese Academy of Sciences\\
    School of Artificial Intelligence, University of Chinese Academy of Sciences\\
    Beijing, China\\
    \{xuzhiwei2019, zhangbin2020, lidapeng2020, zhangzeren2021, zhouguangchong2021, chenhao2019, guoliang.fan\}@ia.ac.cn
}

\begin{document}

\maketitle

\begin{abstract}
Almost all multi-agent reinforcement learning algorithms without communication follow the principle of centralized training with decentralized execution. During the centralized training, agents can be guided by the same signals, such as the global state. However, agents lack the shared signal and choose actions given local observations during execution. Inspired by viewpoint invariance and contrastive learning, we propose consensus learning for cooperative multi-agent reinforcement learning in this study. Although based on local observations, different agents can infer the same consensus in discrete spaces without communication. We feed the inferred one-hot consensus to the network of agents as an explicit input in a decentralized way, thereby fostering their cooperative spirit. With minor model modifications, our suggested framework can be extended to a variety of multi-agent reinforcement learning algorithms. Moreover, we carry out these variants on some fully cooperative tasks and get convincing results.\looseness=-1
\end{abstract}


\section{Introduction}

Multi-agent reinforcement learning has received increasing attention recently due to its successful applications in games~\cite{Berner2019Dota2W, Vinyals2019GrandmasterLI}, transportation~\cite{Chu2020MultiAgentDR, Zhang2019CityFlowAM}, and networks~\cite{Nasir2019MultiAgentDR}. And multi-agent cooperation has gradually gained popularity as a research field: how to control multiple agents to work together to solve the same problem simultaneously? Many multi-agent reinforcement learning algorithms based on centralized training with decentralized execution (CTDE) paradigms~\cite{Lowe2017MultiAgentAF} have been proposed, including multi-agent deep deterministic policy gradient (MADDPG)~\cite{Lowe2017MultiAgentAF}, counterfactual multi-agent policy gradients (COMA)~\cite{Foerster2018CounterfactualMP}, and numerous value decomposition methods~\cite{Sunehag2018ValueDecompositionNF, Rashid2018QMIXMV, Rashid2020WeightedQE, Wang2020ROMAMR, Wang2021RODELR, Mahajan2019MAVENMV}. All of the above methods use a centralized critic to address the instability problem in multi-agent systems.\looseness=-1

Methods built on CTDE, however, have a flaw. While an agent can share the information with other agents during the centralized learning, it can only make decisions regarding its own individual observations for decentralized execution. The lack of common guidance is particularly severe in the partially observable Markov decision process (POMDP)~\cite{Monahan1982StateOT}. As a result, agents lack the desire to cooperate when executing. This issue can be resolved by using certain communication-based multi-agent reinforcement learning techniques~\cite{Sukhbaatar2016LearningMC, Foerster2016LearningTC, Peng2017MultiagentBN, Jiang2018LearningAC, Kim2019LearningTS}. However, communication methods also bring about problems like choosing the information to pass and additional bandwidth requirements.

\begin{figure}[t]
    \centering
    \includegraphics[width=2.7 in]{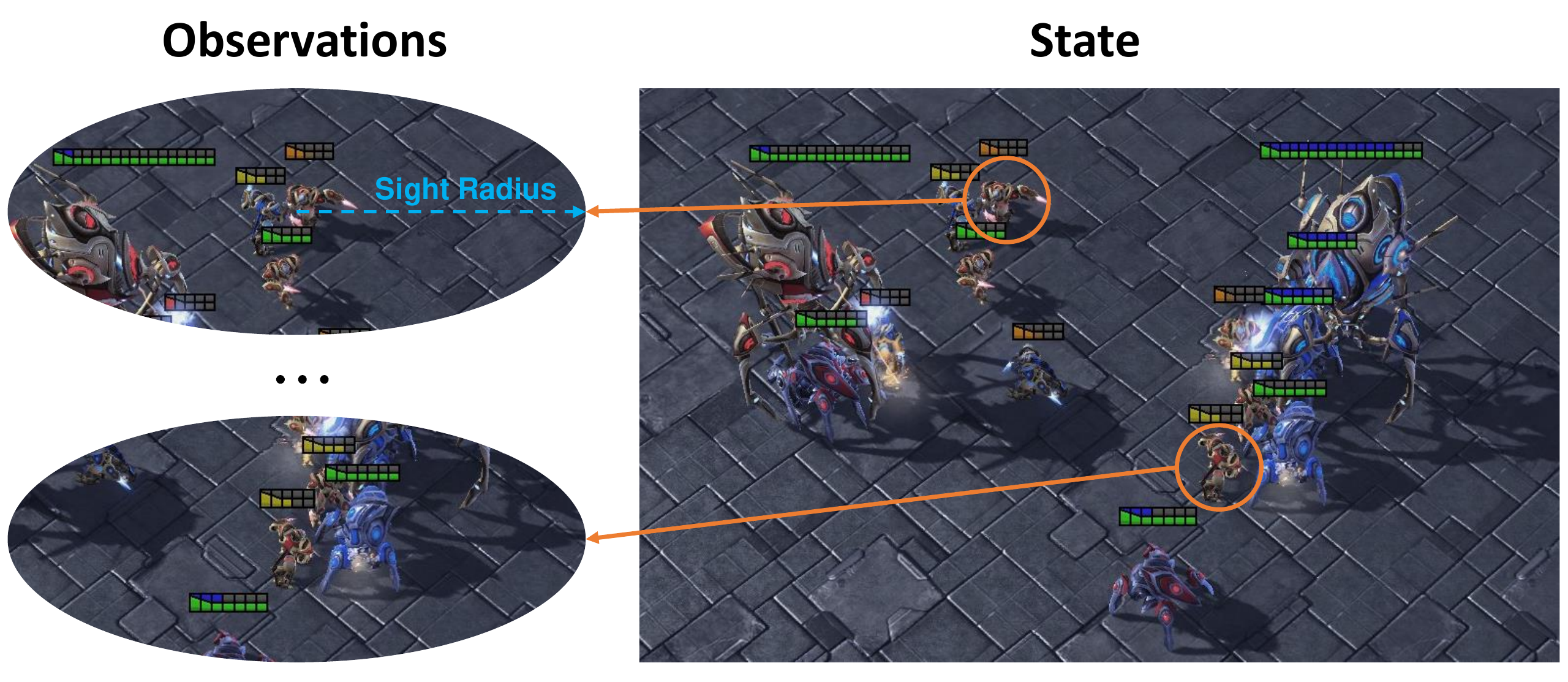}
    \caption{An example on the viewpoint invariance. In the StarCraft multi-agent challenge, agents receive local observations drawn within their field of view. Even though observations of all agents are completely different at each timestep, they nonetheless reflect the same global state.}
    \label{fig:intro}
\end{figure}

Inspired by computer vision~\cite{Morel2009ASIFTAN, Xia2012ViewIH}, we propose \textbf{CO}nsensus \textbf{L}e\textbf{A}rning (COLA) for multi-agent reinforcement learning based on the invariance to views. In POMDPs, despite the fact that local observations of each agent are unique, the observations of all agents at the same moment are different representations of the same global state, as shown in Figure~\ref{fig:intro}. It is like a multi-view problem~\cite{Xu2013ASO}, where the same object can look different from various viewpoints. If the agents are aware of the invariance, they can explicitly select cooperative actions by inferring the same consensus on the state from their different local observations during the decentralized execution. Meanwhile, there is no communication throughout the entire process above.

There are several ways to represent the consensus of the agents on the state. The easiest way is to use the real state as consensus. However, it is not feasible in the implementation to accurately infer the global state from local observations. As a representation learning framework based on the similarity between samples, contrastive learning~\cite{Sermanet2018TimeContrastiveNS, Wu2020OnMI, Khosla2020SupervisedCL, Chen2020ASF, Oord2018RepresentationLW, Chen2020ImprovedBW} has recently emerged as the most popular field in self-supervised learning. While ensuring that the learned representation can retain as much information as possible, contrastive learning promotes similar samples to be close together in the representation space. COLA utilizes contrastive learning to encode local observations into discrete latent spaces through the form of invariances. Even though the local observations between agents at the same moment may be quite dissimilar, we still treat them as similar when training the consensus representation model. Thus, even if agents in COLA receive entirely different local observations, they can still infer the same discrete state latent variables during execution. COLA enables agents to learn the ``tacit agreement" in human social activities: although there is no substantial interaction between agents, they have reached a consensus on the current state.\looseness=-1

We evaluate COLA on variants of the multi-agent particle environments~\cite{Lowe2017MultiAgentAF}, the challenging micromanagement task of StarCraft II~\cite{Samvelyan2019TheSM}, and the mini-scenarios of Google Research Football~\cite{Kurach2020GoogleRF}. We demonstrate that COLA outperforms the previous baselines through experimental results. Furthermore, our approach of simply adding a one-hot consensus encoding to the network input can be extended to any other multi-agent reinforcement learning algorithm.


\section{Related Work}

\subsection{Contrastive Learning}

Many recent studies have focused on mining knowledge from data through self-supervised learning to improve the ability of models to solve downstream tasks. As a typical discriminative method, contrastive learning is easier to train and understand than other generative self-supervised learning methods~\cite{Kingma2014AutoEncodingVB, Goodfellow2014GenerativeAN}. A dataset in contrastive learning tasks is typically organized by similar and dissimilar pairs. Then contrastive learning methods are trained by reducing the distance between augmented views of the same sample and increasing it between different ones~\cite{Wang2020UnderstandingCR}. In this way, contrastive learning enables the representation model to learn the general knowledge contained in the dataset. However, the selection of dissimilar instances, or negative instances, can largely influence the performance of contrastive learning. This process often involves manual selection, limiting further development of contrastive learning. Some new work~\cite{Grill2020BootstrapYO, Chen2021ExploringSS, Caron2021EmergingPI, Zbontar2021BarlowTS, Bardes2021VICRegVR} cancels the explicit introduction of negative examples by using techniques like clustering~\cite{Caron2020UnsupervisedLO}, but they can still get good performances and avoid collapse, a common failure in contrastive learning.

\subsection{Contrastive Learning for Reinforcement Learning}

Since contrastive learning can learn the underlying features of the data, it can be used to extract key information from the raw inputs in reinforcement learning tasks. By comparing states (often in pixel spaces) at different times, contrastive learning can map the original complex state to a low-dimensional latent space, removing redundant information unrelated to learning. So this feature extraction step can improve the sample efficiency of single-agent reinforcement learning, especially in high-dimensional environments. Some studies~\cite{Srinivas2020CURLCU, Kostrikov2021ImageAI, Laskin2020ReinforcementLW, Schwarzer2021DataEfficientRL} successfully use contrastive learning as representation learning before reinforcement learning and achieve state-of-the-art data efficiency on pixel-based RL tasks. Another approach is Contrastive Predictive Coding~\cite{Oord2018RepresentationLW, Hnaff2020DataEfficientIR}, which uses contrastive learning as an auxiliary task to predict the future state conditioned on past observations and actions. In addition, contrastive learning can provide a reward function within a reinforcement learning system~\cite{Sermanet2018TimeContrastiveNS, Dwibedi2018LearningAR}.\looseness=-1

Compared with single-agent reinforcement learning, there is less work in the multi-agent domain for contrastive learning. \citeauthor{Liu2021SocialNC} \shortcite{Liu2021SocialNC} induce the emergence of a common language by maximizing the mutual information between messages of a given trajectory in a contrastive learning manner. \citeauthor{Lo2022LearningTG} \shortcite{Lo2022LearningTG} propose a contrastive method for learning socially-aware motion representations and successfully applied it in human trajectory forecasting and crowd navigation algorithms. Inspired by invariance to views, our proposed COLA can lead agents to reach a tacit agreement in a multi-agent setup where agents only receive partial observations of the same state. To the best of our knowledge, this is the first study to introduce contrastive learning into the decentralized partially observable Markov decision process problem.


\section{Preliminaries}

\subsection{Dec-POMDP and CTDE Methods}

A fully cooperative multi-agent task can be viewed as a decentralized partially observable Markov decision process (Dec-POMDP)~\cite{Oliehoek2016ACI}, which can be modeled as a tuple $G=\langle S, U, A, P, r, Z, O, n, \gamma \rangle$. $s \in S$ is the global state of the environment. In the CTDE principle, the global state $s$ is accessible during the centralized training but not during the decentralized execution. At each time step, each agent $a\in A:=\{1, \dots,n\}$ determines the appropriate action $u^a \in U$ conditioned on local observation $z^a\in Z$, which is provided by the observation function $O(s,a):S \times A \to Z$.  $\boldsymbol{u}\in \boldsymbol{U} \equiv U^n$ denotes the joint action of all agents. The state transition function is written as $P(s^\prime \mid s, \boldsymbol{u}):S\times \boldsymbol{U} \times S \to [0,1]$. All agents in Dec-POMDPs share one reward function: $r(s, \boldsymbol{u}): S\times \boldsymbol{U}\to \mathbb{R}$. The last term $\gamma$ is the discount factor.

Then we introduce some CTDE methods to solve Dec-POMDP problems. A mixing network and several agent networks make up the framework of value decomposition methods. For agent $a$, we can get the individual action-value estimation $Q_a(\tau^a, \cdot)$ by feeding the local observations to its own value net. The mixing network decomposes the joint action-value function into values related to per-agent utilities. To guarantee the consistency between the local greedy actions and global ones, we make the following assumption known as the Individual-Global-Max (IGM)~\cite{son2019qtran}:
\begin{equation*}
    \arg \max _{u^a} Q_{tot}(\boldsymbol{\tau}, \boldsymbol{u})=\arg \max _{u^a} Q_{a}\left(\tau^{a}, u^{a}\right), \quad \forall a \in A,
\end{equation*}
where $\boldsymbol{\tau}$ represents the joint action-observation histories. Value decomposition methods are trained end-to-end by minimizing the loss function:
\begin{equation}
\mathcal{L}=\left(y_{tot}-Q_{tot}(\boldsymbol{\tau}, \boldsymbol{u})\right)^2,
\label{eq:tderror}
\end{equation}
where $y_{tot}=r+\gamma \max_{\boldsymbol{u}^\prime} \hat{Q}_{tot}(\boldsymbol{\tau}^\prime,\boldsymbol{u}^\prime )$. $\hat{Q}_{tot}(\cdot)$ is the target network of the joint action-value function. In the value decomposition method, the agent greedily chooses actions according to its individual action-value function $Q_a$ during the decentralized execution.\looseness=-1

MADDPG~\cite{Lowe2017MultiAgentAF} is an actor-critic method based on the CTDE paradigm for continuous action tasks. Each agent learns a deterministic policy $\mu_a$ in an off-policy manner. MADDPG consists of a centralized critic and several independent actors corresponding to each agent. During the centralized training phase, the critic can use global information to estimate the joint action-value  $Q_{a}\left(s, u^{1}, \ldots, u^{n}\right)$ for each agent $a$. We can train the critic by minimizing the TD-error like Eq.~\eqref{eq:tderror}. The gradient of the individual deterministic policy $\mu_a$ can be derived as follow:
\begin{equation}
    \begin{split}
    \nabla &J\left(\mu_{a}\right)=\\&\mathbb{E}_{\mathcal{D}}\left[\left.\nabla \mu_{a}\left(\tau^{a}\right) \nabla_{u^{a}} Q_{a}\left(s, u^{1}, \ldots, u^{n}\right)\right|_{u^{a}=\mu_{a}\left(\tau^{a}\right)}\right],
    \end{split}
\end{equation}
where $u^a$ in $Q_{a}\left(s, u^{1}, \ldots, u^{n}\right)$ is sampled from the current policy $\mu_a$, while the actions of other agents are sampled from the replay buffer $\mathcal{D}$.

Value decomposition methods and policy-based multi-agent methods are very different in structure. However, they both follow CTDE principles, and suffer from a lack of common guidance during the decentralized execution in fully cooperative tasks.

\subsection{Knowledge Distillation with No Labels}

Many early contrastive learning studies require a large batch pool or suitable negative samples, so they cannot be directly applied to reinforcement learning. In Knowledge Distillation with No Labels (DINO)~\cite{Caron2021EmergingPI} which is inspired by knowledge distillation~\cite{Hinton2015DistillingTK}, a student network $g_{\theta_S}$ directly predicts the output of a teacher network $g_{\theta_T}$ to simplify self-supervised training. For two different augmentations of the same image, $x$ and $x^\prime$, two networks output probability distributions over K dimensions, represented as $P_T(x)$ and $P_S(x^\prime)$ respectively. $P_T(x)$ is given as follows:
\begin{equation}
P_{T}(x)^{(i)}=\frac{\exp \left(g_{\theta_{T}}(x)^{(i)} / \tau_{T}\right)}{\sum_{k=1}^{K} \exp \left(g_{\theta_{T}}(x)^{(k)} / \tau_{T}\right)}, \forall i \in \{1, 2,\dots,K\}.
\label{eq:ce_loss}
\end{equation}
$\tau_T$ is the temperature coefficient that controls the sharpness of the output dimension. A similar formula holds for $P_S(x^\prime)$ with temperature $\tau_s$. So in the case where no labels are available, the teacher network can give pseudo-labels to unlabelled data. The student network needs to match these pseudo-labels by minimizing the cross-entropy loss:
\begin{equation*}
\min _{\theta_S}\sum_{\substack{x,x^\prime\in \mathcal{V} \\ x \neq x^\prime}} H(P_T(x),P_S(x^\prime)),
\end{equation*}
where $\mathcal{V}$ represents different views of an image and $H(a,b)$ $= -a\log b$. Unlike general knowledge distillation methods, the structure of the teacher network in DINO is consistent with that of the student network, and its weight parameters also use an exponential moving average (EMA) on the student weights. Furthermore, to avoid trivial solutions, the centering operation prevents one dimension from dominating but encourages the output of the teacher network to the uniform distribution. So we obtain the modified output of the teacher network according to Eq.~\eqref{eq:ce_loss}:
\begin{equation}
P_{T}(x)= \operatorname{Softmax}\left((g_{\theta_{T}}(x) - \lambda) / \tau_{T}\right),
\label{eq:modified_ce_loss}
\end{equation}
where the center $\lambda$ is updated with an exponential moving average of $g_{\theta_{T}}(x)$. DINO applies the above framework to the Vision Transformer and obtains features that outperform those trained with labels.

In multi-agent systems, observations of each agent can be seen as different views of the global state. Agents in COLA can learn the identical discrete representation from these different local observations by employing contrastive learning. We believe the inferred consensus can ignite collaboration between agents during the decentralized execution.


\section{Consensus Learning}

In this section, we propose COLA, a novel multi-agent cooperative reinforcement learning framework that can explicitly guide agents to make cooperative decisions in decentralized execution.
It is worth noting that our proposed method is universal and any CTDE multi-agent algorithm already in existence can be combined with COLA.

\begin{figure}[tb]
    \centering
    \includegraphics[width=2.7 in]{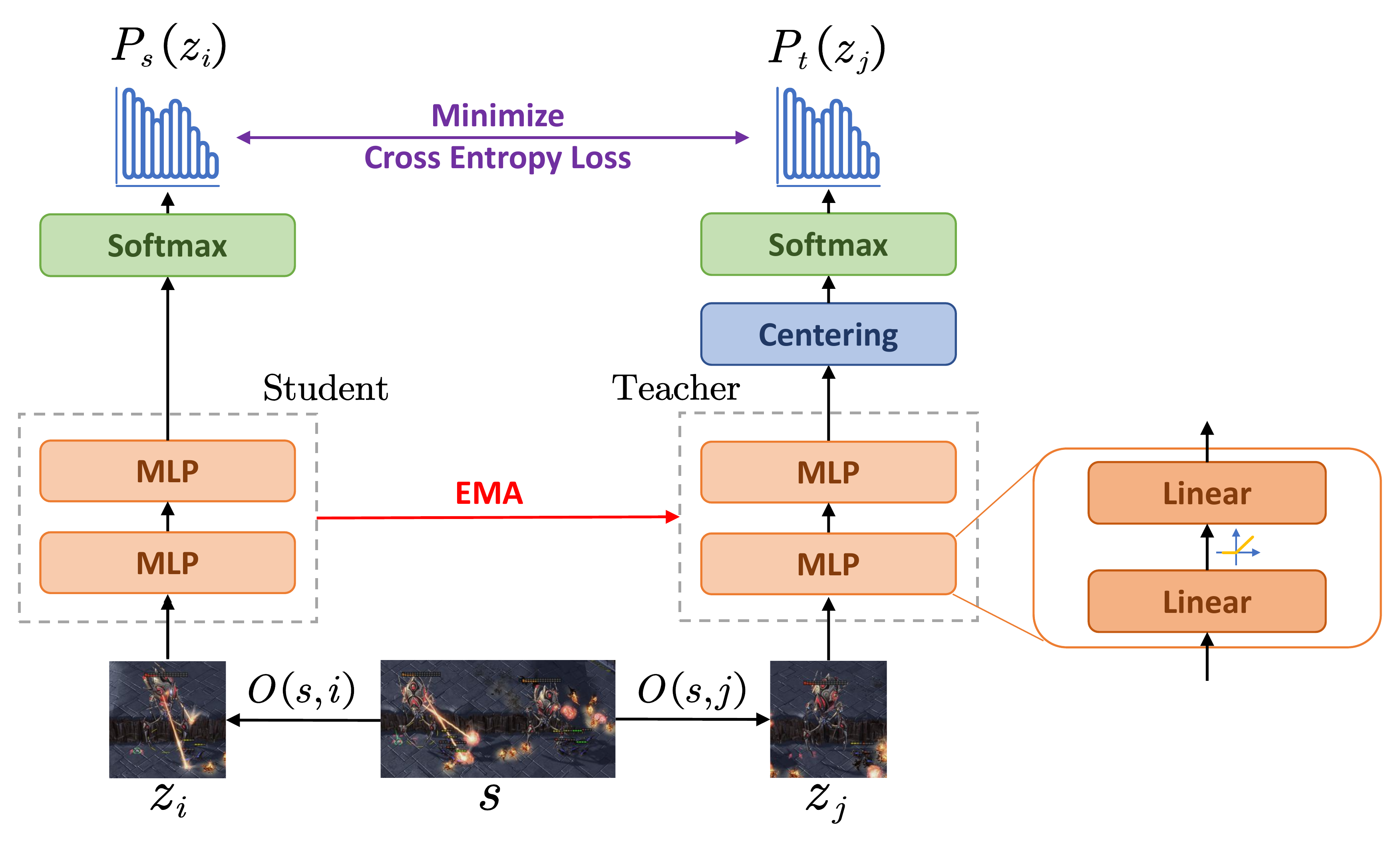}
    \caption{The overall architecture of the consensus builder.}
    \label{fig:consensus_builder}
\end{figure}

\subsection{Consensus Builder}

During the centralized training phase of the conventional CTDE methodology, each agent can improve coordination based on consistent environmental signals, such as the same global state. Even in decentralized execution, agents also need to derive shared signals based on their local information to get consistent guidance and inspiration. In a multi-agent system, while local observations are different for each agent, they all correspond to the same global state. This is similar to the multi-view problem in computer vision.
Data augmentation~\cite{Shorten2019ASO} is a method that enables limited data to generate more equivalent data, hence increasing the diversity of training data sets.
We can treat observation function $O(s,a)\in Z$ of each agent as an augmentation operation. The global state $s$ is data-augmented to form local observation $z$ of each agent. However, it is not feasible to directly map local observations to a continuous low-dimensional space to obtain shared signals using general contrastive learning methods, because there are always subtle differences between the continuous signals inferred by different agents. Furthermore, since both consensus learning and reinforcement learning are carried out simultaneously in COLA, the best way to improve robustness and make learning easier is to map the consensus to a discrete space. To achieve this goal, we adopt a DINO-like framework to learn discrete consensus, named consensus builder. The goal of the consensus builder is to group local observations for the same state into the same class $c$. Then the class $c$ is the discrete global consensus shared by all agents in that state.\looseness=-1

Enlightened by DINO, we enable the consensus builder to simulate knowledge distillation through student and teacher networks as shown in Figure~\ref{fig:consensus_builder}. The centering operation on the output of the teacher network avoids collapse. In multi-agent reinforcement learning, we naturally treat the observation function as data augmentation operations. In addition, since most of the observations in the multi-agent environments we used in this paper are low-dimensional and not complex high-dimensional signals like images, simple multi-layer perceptrons (MLP) are enough to be used to build the main body of the network. For the state $s$, $z^a=O(s,a)\in Z$ represents the local observation of the agent $a$. According to Eq.~\eqref{eq:ce_loss} and Eq.~\eqref{eq:modified_ce_loss}, we can obtain the probabilities $P_T(z^a)$ and $P_S(z^a)$ of the agent's discrete consensus under the condition of local observation $z$, where $a\in \{1, 2, \dots, n\}$. We perform the pairwise comparison between the inferred consensus of all agents and minimize the cross-entropy loss:
\begin{equation}
\mathcal{L}_{CB}=\sum_{a, b} H(P_T(z^a),P_S(z^b)), 
\label{eq:cb_loss}
\end{equation}
where $a, b \in \{1,2,\dots n\}$ and $a\neq b$. 
By minimizing $\mathcal{L}_{CB}$, consensus builders can provide identical categorical distributions for different views of the same state.
$c^a\in \{1, 2,\dots,K\}$ represents the discrete consensus inferred by the agent $a$ based on the observation $z^a$, which can be computed according to the output of the student network:
\begin{equation}
c^a=\arg\max_c P_S(z^a)^{(c)}.
\label{eq:consensus}
\end{equation}
Since Eq.~\eqref{eq:cb_loss} requires local observations of all agents and Eq.~\eqref{eq:consensus} only requires that of a single agent, the consensus builder also follows the paradigm of centralized training with decentralized execution. The center $\lambda$ in Eq.~\eqref{eq:modified_ce_loss} and the teacher network are updated with an exponential moving average. All hyperparameter settings and the details of the implementation can be found in Appendix B.

\begin{figure}[t]
    \centering
    \subfigure{
    \includegraphics[width=1.2 in]{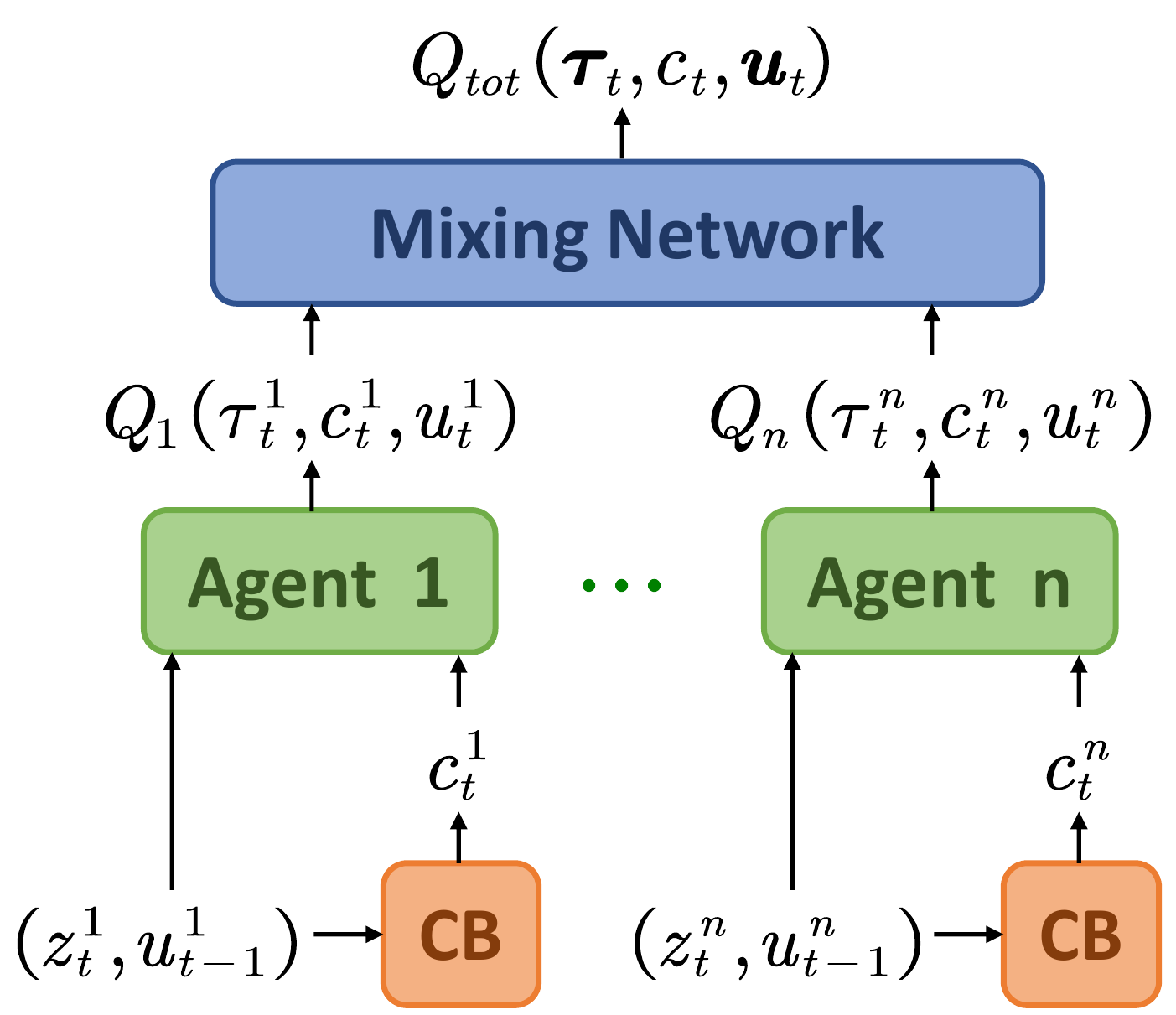}
    \label{fig:vdn}
    }
    \hspace{0.1 in}
    \subfigure{
    \includegraphics[width=1.5 in]{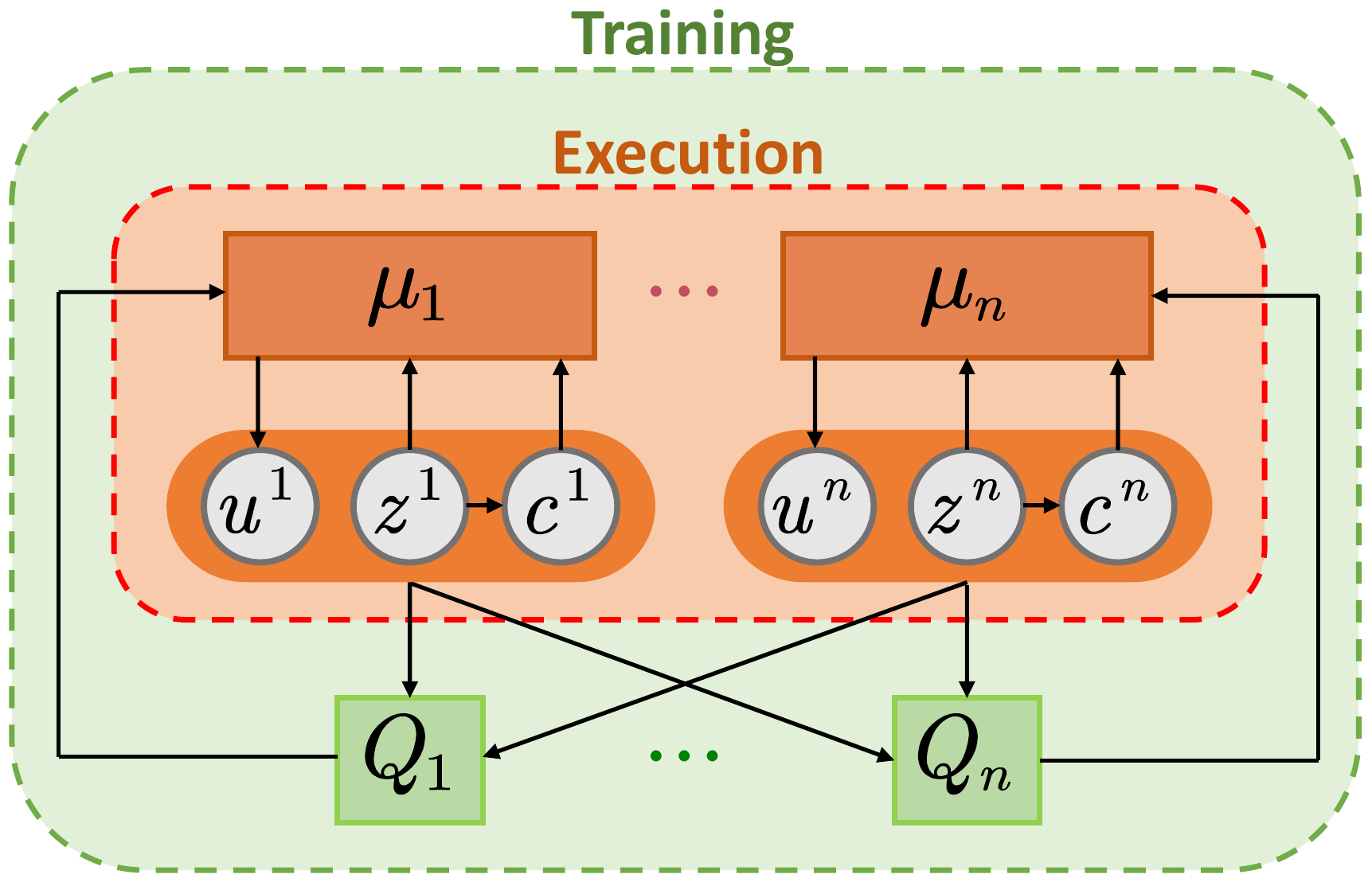}
    \label{fig:maddpg}
    }
\caption{Illustration of multi-agent reinforcement learning methods with consensus learning. \textbf{Left:} The value decomposition method with consensus learning. \textbf{Right:} The multi-agent actor-critic method with consensus learning.}
\label{fig:implementation}
\end{figure}

\subsection{COLA Architecture}

\begin{figure*}[t]
    \centering
    \includegraphics[width=5.7 in]{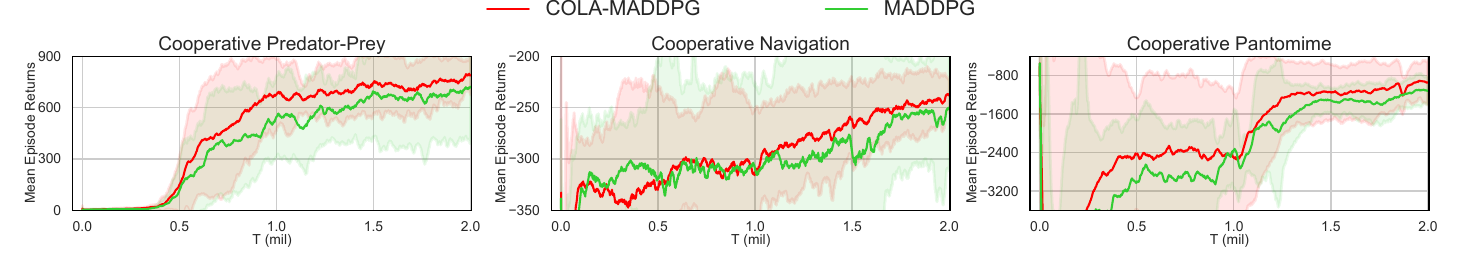}
    \caption{Mean episode return on different tasks in the multi-agent particle environment.}
    \label{fig:mpe_result}
\end{figure*}

Because the consensus builder can judge the current consensus in decentralized execution, we can use it as additional information for each agent's decision-making. We concatenate local observations with the one-hot consensus and feed them into the network. This simple method can be easily deployed to various multi-agent reinforcement learning algorithms. We input the merged information into the agent network so that value decomposition methods can be trained end-to-end to minimize the following loss:
\begin{equation}
\mathcal{L}_{RL}=\left(y_{tot}-Q_{tot}(\boldsymbol{\tau}, \boldsymbol{c}, \boldsymbol{u})\right)^2,
\label{eq:ctd_loss}
\end{equation}
where $y_{tot}=r+\gamma \max_{\boldsymbol{u}^\prime} \hat{Q}_{tot}(\boldsymbol{\tau}^\prime,\boldsymbol{c}^\prime,\boldsymbol{u}^\prime )$. While for policy-based multi-agent reinforcement learning methods such as MADDPG, $Q_{a}\left(s, c^{a}, u^{1}, \ldots, u^{n}\right)$ denotes the centralized action-value function of agent $a$. And the following policy gradient can be calculated:
\begin{equation}
\begin{split}
&\nabla J\left(\mu_{a}\right)=\\
&\mathbb{E}_{\mathcal{D}}\left[\left.\nabla \mu_{a}\left(\tau^{a},c^{a}\right) \nabla_{u^{a}} Q_{a}\left(s, c^{a}, u^{1}, \ldots, u^{n}\right)\right|_{u^{a}=\mu_{a}\left(\tau^{a},c^{a} \right)}\right].
\end{split}
\label{eq:policy_gradient}
\end{equation}
Figure~\ref{fig:implementation} depicts the overview of the aforementioned multi-agent reinforcement learning methods with consensus learning. Since the inferred consensus is discrete, we do not propagate gradients about reinforcement learning through the consensus builder. As a result, the complexity of the reinforcement learning model was not significantly increased. So if COLA can improve the performance, we can clearly know that the improvement is brought by the consensus learning mechanism rather than by increasing model capacity. To prevent the sparse input due to the large number of consensus classes $K$, we add an embedding layer to map the originally one-hot hard consensus signal into a dense low-dimensional space. Although different local observations guide the agent to make different decisions, the same consensus given by the consensus builder based on invariance drives the collaboration of the agents. We argue that consensus learning can solve the problem of agents lacking cooperative signals when they execute in a decentralized way.\looseness=-1

\begin{figure*}[t]
    \centering
    \includegraphics[width=5.7 in]{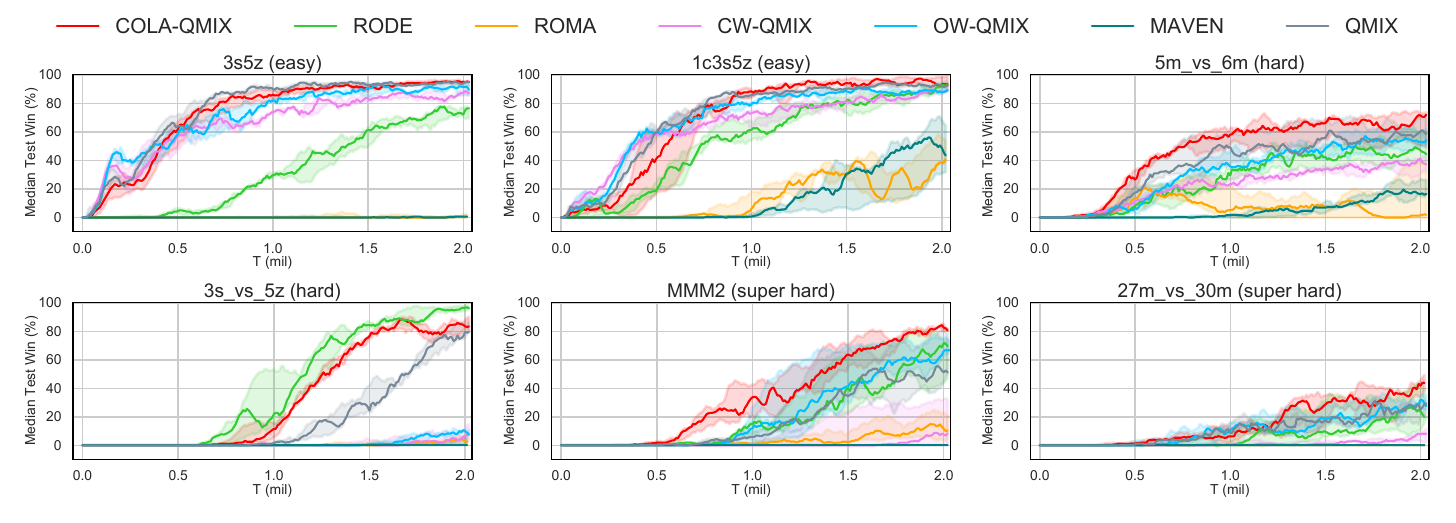}
    \caption{Performance comparison with baselines in different SMAC scenarios.}
    \label{fig:smac_result}
\end{figure*}


\section{Experiment}

We evaluate COLA in three challenging environments: the multi-agent particle environments (MPE), the StarCraft multi-agent challenge (SMAC), and Google Research Football (GRF). The detailed descriptions for the three environments can be found in Appendix A. We apply COLA to different CTDE algorithms for different environments and then compare them with baselines. Finally, we visualize the consensus inferred by agents. In this section, we present experimental results and answer the following three questions: (a) Can COLA be applied to various CTDE algorithms and improve their performance? (b) Is the superior performance of COLA due to consensus learning? (c) How does the hyperparameter $K$ affect the performance of COLA?

\subsection{Performance on Multi-agent Particle Environments}

There are different continuous action tasks in the multi-agent particle environments. We selected three representative cooperation scenarios: \emph{Cooperative Predator-Prey}, \emph{Cooperative Navigation}, and a new scenario we designed, \emph{Cooperative Pantomime}. In \emph{Cooperative Predator-Prey}, a single prey is the built-in heuristic agent, and we control a team of three predators. \emph{Cooperative Pantomime} is similar to \emph{Cooperative Communication} without the communication mechanism. The two agents in \emph{Cooperative Pantomime} want to get to their target landmark, which is only known by another agent. So the agent can only infer the location of its goal from actions of another agent. Details of the environment can be found in Appendix A.1.

Since MADDPG often fails in complex environments like SMAC, we compare MADDPG with COLA-MADDPG in multi-agent particle environments, where COLA-MADDPG is obtained by combining the consensus builder with MADDPG. We carry out each experiment with 5 random seeds, and the mean episode returns are shown with a 95\% confidence interval. Since the scenarios in MPE is simple, we set the number of classes of consensus $K=4$.

Figure~\ref{fig:mpe_result} shows the learning curves of COLA-MADDPG and MADDPG in different scenarios. COLA-MADDPG can successfully solve the task in any scenario and performs better than MADDPG. Each agent in MADDPG corresponds to a policy network, which may not be conducive to encouraging agents to cooperate even in the same state. In COLA-MADDPG, we feed the signals inferred by the consensus builder as explicit input to each policy network, which helps the agents to have an abstract understanding of the state.

\begin{figure*}[t]
    \centering
    \includegraphics[width=5.7 in]{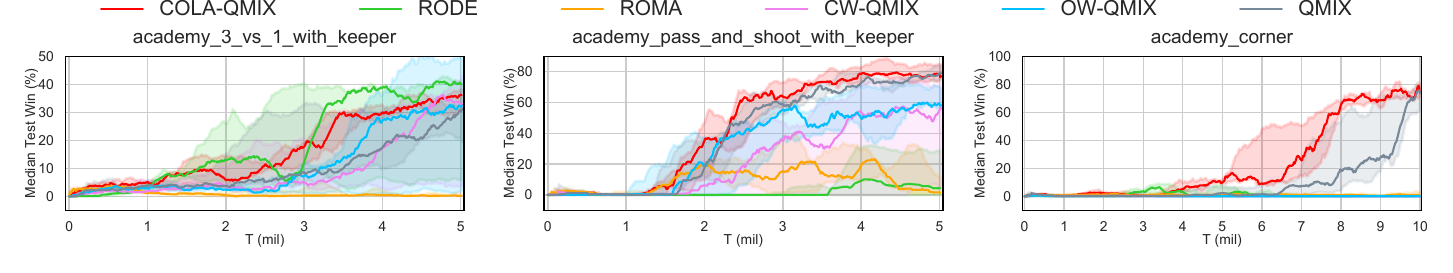}
    \caption{Comparison of our approach against baselines on Google Research Football.}
    \label{fig:grf_result}
\end{figure*}

\begin{figure}[t]
    \centering
    \includegraphics[width=3.0 in]{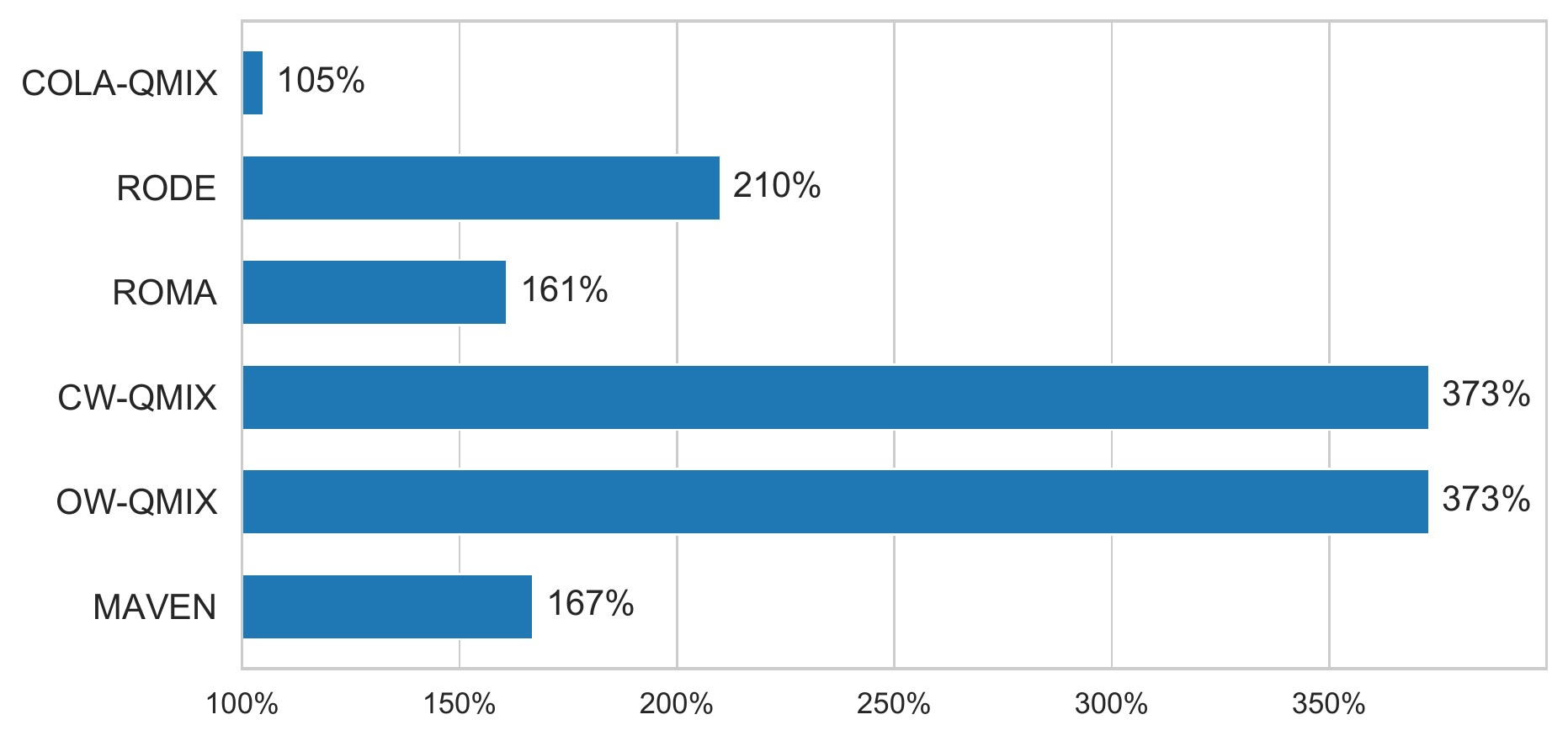}
    \caption{The bars represent the relative size of neural network models related to reinforcement learning for some value decomposition variants over QMIX.}
    \label{fig:model_size}
\end{figure}

\subsection{Performance on StarCraft II and Google Research Football}

Many value decomposition methods have achieved impressive performance in complex scenarios such as StarCraft II and Google Research Football. We develop COLA-QMIX by combining COLA with the vanilla value decomposition method QMIX~\cite{Rashid2018QMIXMV}. To demonstrate the superiority of our proposed COLA framework, we evaluate COLA-QMIX and baselines including QMIX, MAVEN~\cite{Mahajan2019MAVENMV}, ROMA~\cite{Wang2020ROMAMR}, Weighted QMIX~\cite{Rashid2020WeightedQE}, and RODE~\cite{Wang2021RODELR} on SMAC and Google Research Football. To eliminate unfair comparisons like those present in the original literature of some baselines, we evaluate the algorithms in the exact same environmental settings and show their real performance.

SMAC is a novel multi-agent testbed containing various micro-management tasks and we selected several representative scenarios. The version of StarCraft we used is SC2.4.6.2.69232 instead of the simpler SC2.4.10. The multi-agent cooperation problems in hard and super hard scenarios have a large state space, so we set $K=32$. For the easy scenarios, however, we still keep $K=4$. Furthermore, the number of surviving agents always changes in one episode. Local observations of dead agents are padded with zeros, which violates the viewpoint invariance principle. Therefore, we disregard local observations of dead agents when training the consensus builder.

The median win ratios are shown in Figure~\ref{fig:smac_result}, and the 25-75\% percentiles are shaded. In the homogeneous agent scenarios \emph{5m\_vs\_6m} and \emph{27m\_vs\_30m}, COLA-QMIX improves the performance of vanilla QMIX and outperforms other methods. The \emph{MMM2} scenario contains a wide variety of agents. COLA-QMIX can map the observations of different types of agents into a low-dimensional space so that the agents get the same cooperative signal. Besides, COLA still achieves slight performance improvement on easy scenarios, which is difficult for complex value decomposition variants. We also compare the performance of COLA-VDN and VDN in SMAC without access to the global state. The results in Appendix C show that the consensus inferred by COLA can indeed be used as a global shared signal like the state.

In Google Research Football, agents can be trained to play football. There are many mini-scenarios in Football Academy. To speed up the training, we take the goal of our players or the ball into our half as the basis for determining the termination of an episode. In addition to scoring rewards, the agents can receive rewards when the ball approaches the goal. Furthermore, there are many types of agents, including strikers and midfielders. We performed COLA-QMIX and other baselines in three scenarios on Google Research Football.\looseness=-1

Figure~\ref{fig:grf_result} shows the learning curves of different algorithms in Google Research Football. We focus on comparing the performance of vanilla QMIX and its variant with consensus learning. \emph{Academy\_corner} is a challenging scenario where players require more coordination to score goals. The other players need to agree with the server in the corner to reduce the difficulty of scoring. COLA-QMIX outperforms other algorithms in almost all scenarios, especially in \emph{academy\_corner}. 

In summary, we apply COLA to the multi-agent actor-critic algorithm and the value decomposition method, respectively. The resulting algorithms achieve convincing results on three different experimental platforms.

\begin{figure*}[t]
    \centering
    \includegraphics[width=6.3 in]{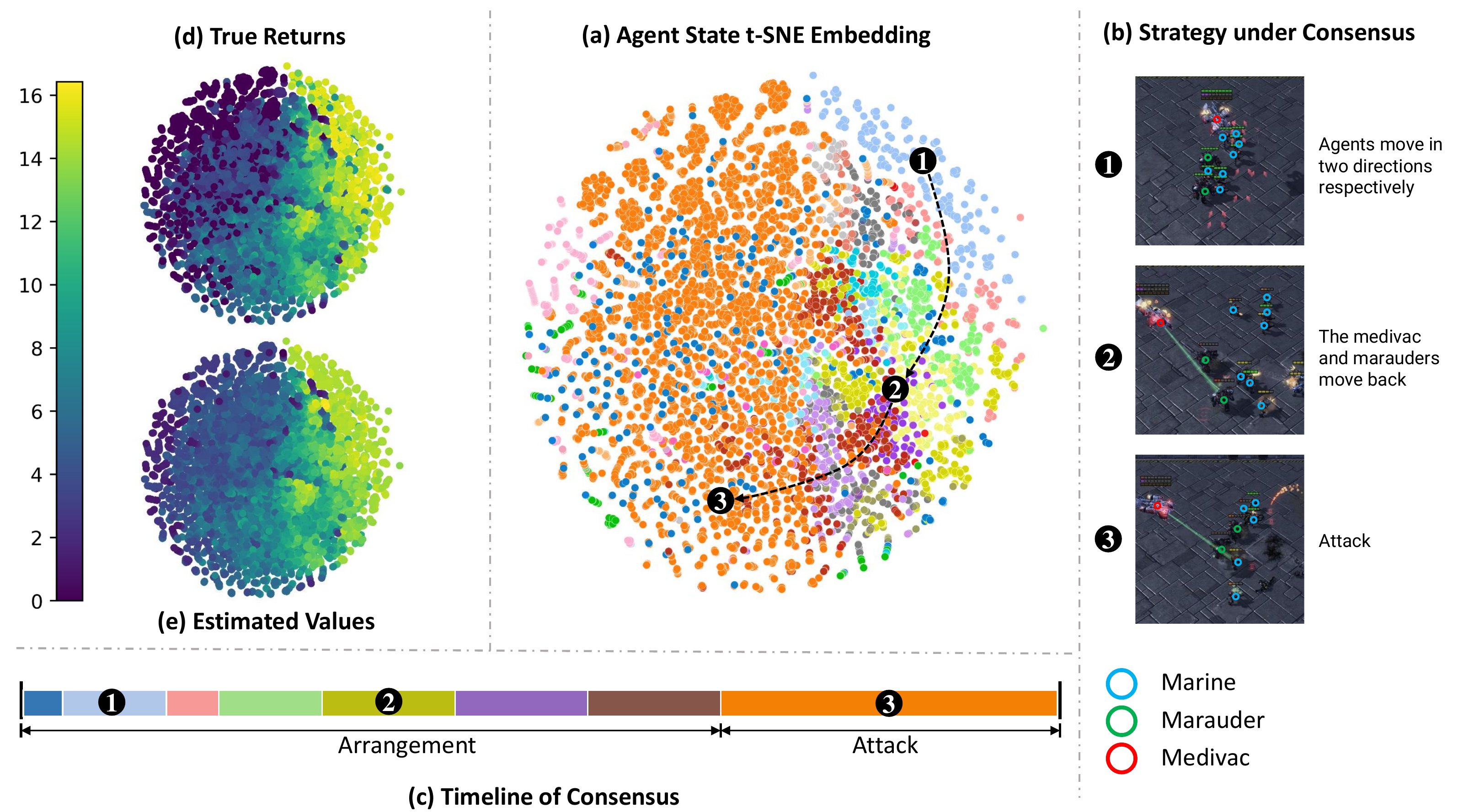}
    \caption{The analysis and visualization of agents' consensus in \emph{MMM2} scenario. (a) The 2D t-SNE embedding of states. Each point is colored according to the consensus inferred by the agents based on local observations in this state. (b) The global strategies of the agents under the guidance of different consensus. (c) The change of consensus in an episode. The consensus represented by each color is consistent with that in (a). (d) The true returns of agents and (e) the estimated values.}
    \label{fig:visualization}
\end{figure*}

\subsection{Case Study and Visualization}

To explore what makes the COLA framework superior to the original algorithm, we first rule out the factor of increasing the capacity of the neural network model. Figure~\ref{fig:model_size} shows the relative size of the reinforcement learning model for various value decomposition variants over QMIX. Since the gradient of reinforcement learning tasks in the COLA framework cannot be transferred to the consensus builder, the actual model size does not increase significantly. Some complex value decomposition variants significantly increase the capacity of the model. The experimental results, however, show that these variants do not result in an improvement proportionate to the size of their models. The COLA framework brings the most notable performance gains with the smallest model.

Then we provide a case study to explore the role of the consensus builder and visualize its impact in an intuitive form. We perform our research on the \emph{MMM2} scenario from the SMAC benchmark. As shown in Figure~\ref{fig:visualization}, in the 2D t-SNE~\cite{Maaten2008VisualizingDU} embedding of states, the states with the same consensus generated by the agent tend to cluster together. Besides, states with the same consensus often have similar estimated values or true returns. It should be noted that the consensus builder learns the discrete consensus only based on the given observations, and the state information is not involved in training. So we can conclude that although the consensus output by the consensus builder is directly determined by the observations of each agent, it can reflect the global state. In other words, the consensus inferred by each agent during the decentralized execution can be used as a shared cooperation signal similar to the global state. Furthermore, when sharing different consensuses, the agents implement different global strategies. By analyzing and visualizing consensuses, we can point out that the consensus signal inferred by the consensus builder is beneficial for agents to cooperate explicitly during the decentralized execution.\looseness=-1

\subsection{Ablation Study}

\begin{figure}[t]
    \centering
    \includegraphics[width=2.1 in]{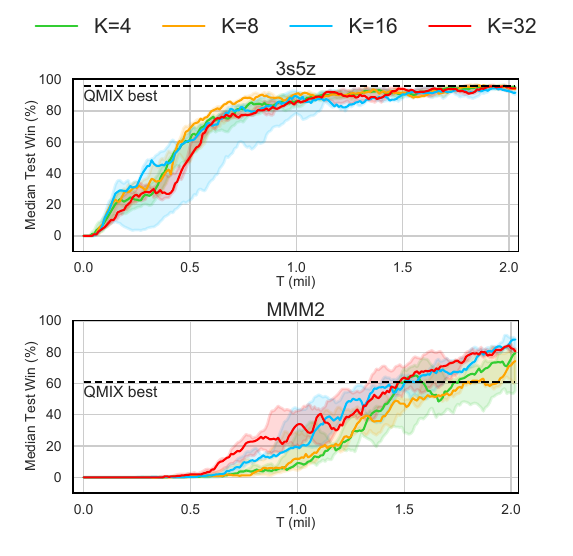}
    \caption{Influence of the $K$ for COLA-QMIX.}
    \label{fig:different_k}
\end{figure}

In this section, we will investigate how the amount of consensus $K$ affects the performance of COLA. We conduct an ablation study on two scenarios with various difficulties in SMAC. While keeping other settings unchanged, we compare the performance of COLA-QMIX under different values of $K$. Figure~\ref{fig:different_k} displays the different results in the easy scenario \emph{3s5z} and the super hard scenario \emph{MMM2}. In the \emph{3s5z} scenario, COLA-QMIX performs the worst when $K=32$ but learns faster when $K$ takes relatively tiny values. In the \emph{MMM2} scenario, conversely, COLA-QMIX performs significantly better when $K=32$ or $K=16$ than when $K$ is small. Therefore, we conclude that the choice of $K$ depends on the difficulty of the task. We should pick a higher $K$ value when the reinforcement learning task is complex and vice versa.

\section{Conclusion}

Inspired by viewpoint invariance, We propose a consensus learning framework called COLA. COLA can infer a shared consensus based on local observations of agents to alleviate the lack of the same guidance during the decentralized execution. We believe that the same consensus can encourage agents to act cooperatively. To the best of our knowledge, COLA is the first work to address the Dec-POMDP problem by integrating contrastive learning into multi-agent reinforcement learning. We combine consensus learning with various multi-agent algorithms and evaluate their performances in several environments. Experimental results demonstrate that our proposed COLA can improve the performance of the original algorithms on fully cooperative tasks. We believe the COLA framework is the most cost-effective, bringing remarkable performance improvement with minor changes of reinforcement learning models. Further research on how to choose the hyperparameter $K$ would be of interest.


\bibliography{main.bbl}

\onecolumn
\appendix
\numberwithin{equation}{section}
\numberwithin{figure}{section}
\numberwithin{table}{section}
\renewcommand{\thesection}{{\Alph{section}}}
\renewcommand{\thesubsection}{\Alph{section}.\arabic{subsection}}
\renewcommand{\thesubsubsection}{\Roman{section}.\arabic{subsection}.\arabic{subsubsection}}
\setcounter{secnumdepth}{-1}
\setcounter{secnumdepth}{3}

\section{Environment Details}

\subsection{Multi-Agent Particle Environments}

\begin{figure*}[h]
    \centering
    \includegraphics[width=5.0 in]{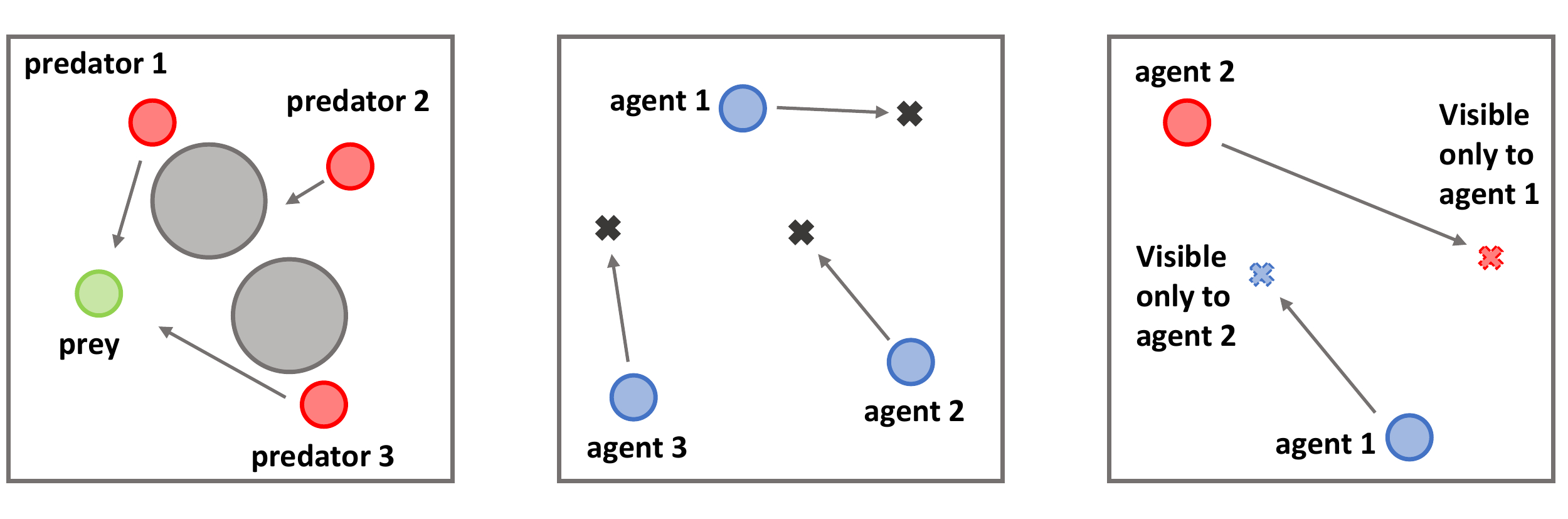}
    \caption{Illustrations of the multi-agent particle environments. \textbf{Left:} \emph{Cooperative Predictor-Prey}. \textbf{Middle:} \emph{Cooperative Navigation}. \textbf{Right:} \emph{Cooperative Pantomime}.}
    \label{fig:mpe}
\end{figure*}

All multi-agent particle environments we tested include \emph{Cooperative Predictor-Prey}, \emph{Cooperative Navigation}, and a scenario we proposed, \emph{Cooperative Pantomime}. The three scenarios are illustrated in Figure~\ref{fig:mpe}. We configure the number of steps in an episode to be 100. Each scenario we tested is described in detail below.

\noindent{\textbf{Cooperative Predator-Prey.}}\quad Prey (green) is controlled by the heuristic algorithm. Predators (red) aim to catch prey that moves faster than them through cooperation. Two landmarks (gray) will be randomly generated in each round to hinder the movement of agents.

\noindent{\textbf{Cooperative Navigation.}}\quad Agents (blue) have to learn to cover all the targets while avoiding collisions, which requires the agents to assign their targets collaboratively.

\noindent{\textbf{Cooperative Pantomime.}}\quad This scenario is similar to the \emph{Cooperative Communication} scenario. Each agent cannot see its own target but can see the target of the other. Therefore, each agent needs the help of the other to be able to approach its goal. Unlike the \emph{Cooperative Communication} task, the agents in \emph{Cooperative Pantomime} cannot communicate with each other. Each agent can only infer the intentions of others from their positions or actions. Since the sight of each agent is different, we believe that consensus learning can improve the cooperation ability of the agents in this case.

\subsection{SMAC}

SMAC is a multi-agent experimental platform with rich micro-management tasks based on the real-time strategy game StarCraft II. The goal of each task is to control different types of agents to move or attack to defeat the enemies. The version of StarCraft II we used in this paper is 4.6.2.69232 rather than the easier version 4.10. When we ran baselines, we used the same environment settings (same version of the StarCraft and same SMAC) for the fairest comparison. The hyperparameters of all baselines are consistent with the official codes. Table~\ref{table:scenario} presents the details of several representative scenarios in SMAC.

\begin{table}[h]
\centering
\begin{tabular}{lcccc}
\hline
\textbf{Name}  & \textbf{Ally Units}                                                         & \textbf{Enemy Units}                                                        & \textbf{Type}                                                                         & \textbf{Difficulty} \\ \hline
3s5z           & \begin{tabular}[c]{@{}c@{}}3 Stalkers\\ 5 Zealots\end{tabular}              & \begin{tabular}[c]{@{}c@{}}3 Stalkers\\ 5 Zealots\end{tabular}              & \begin{tabular}[c]{@{}c@{}}Heterogeneous\\ Symmetric\end{tabular}                     & Easy       \\ \hline
1c3s5z         & \begin{tabular}[c]{@{}c@{}}1 Colossus\\ 3 Stalkers\\ 5 Zealots\end{tabular} & \begin{tabular}[c]{@{}c@{}}1 Colossus\\ 3 Stalkers\\ 5 Zealots\end{tabular} & \begin{tabular}[c]{@{}c@{}}Heterogeneous\\ Symmetric\end{tabular}                     & Easy       \\ \hline
5m\_vs\_6m     & 5 Marines                                                                   & 6 Marines                                                                   & \begin{tabular}[c]{@{}c@{}}Homogeneous\\ Asymmetric\end{tabular}                      & hard       \\ \hline
3s\_vs\_5z     & 3 Stalkers                                                                  & 5 Zealots                                                                   & \begin{tabular}[c]{@{}c@{}}Homogeneous\\ Asymmetric\end{tabular}                      & hard       \\ \hline
MMM2           & \begin{tabular}[c]{@{}c@{}}1 Medivac\\ 2 Marauders\\ 7 Marines\end{tabular} & \begin{tabular}[c]{@{}c@{}}1 Medivac\\ 3 Marauder\\ 8 Marines\end{tabular}  & \begin{tabular}[c]{@{}c@{}}Heterogeneous\\ Asymmetric\\ Macro tactics\end{tabular}    & Super Hard \\ \hline
27m\_vs\_30m   & 27 Marines                                                                  & 30 Marines                                                                  & \begin{tabular}[c]{@{}c@{}}Homogeneous\\ Asymmetric\\ Massive Agents\end{tabular}     & Super Hard \\ \hline
\end{tabular}
\caption{Maps in different scenarios.}
\label{table:scenario}
\end{table}

\subsection{Google Research Football}

In this paper, we choose three official scenarios in Football Academy, which is a collection of some mini-scenarios. \emph{Academy\_3\_vs\_1\_with\_keeper} and \emph{academy\_pass\_and\_shoot\_with\_keeper} are easy scenarios, and \emph{academy\_corner} is the hard one. Agents are rewarded when they score a goal. In addition, when the ball is close to the goal of opponent, our agents can also get rewards. We limit the football to the half of opponent to speed up training. The round ends once the ball enters our half. Observations of the agent include the relative positions of all other entities. Therefore, we believe that establishing consensus among agents based on viewpoint invariance can promote cooperation between agents.

\begin{figure*}[h]
    \centering
    \includegraphics[width=5.4 in]{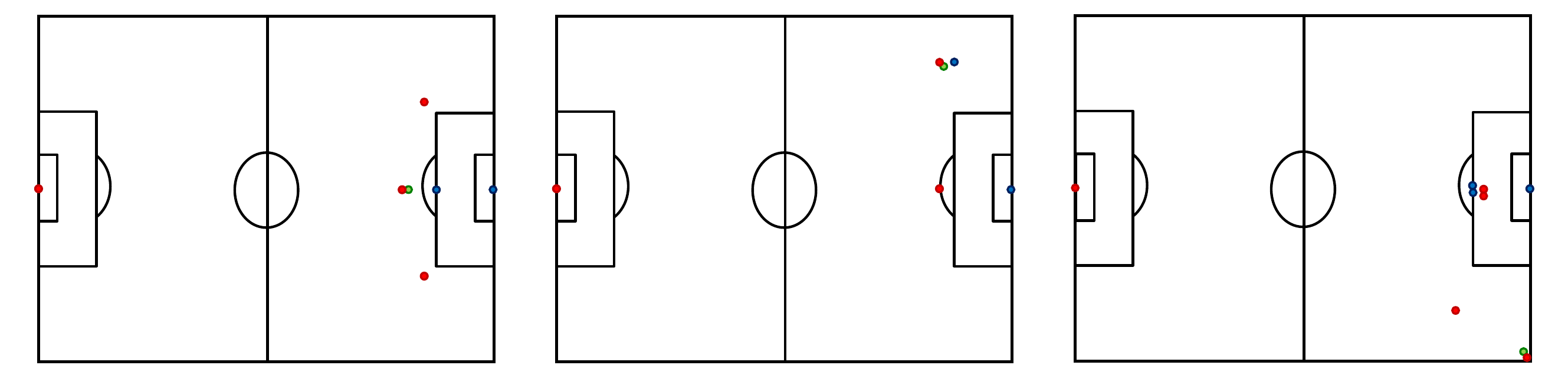}
    \caption{The initial position of each agent in the Google Research Football environments. Our players are represented by the red dots, and the opponents are represented by the blue dots. Green dots represent the football. \textbf{Left:} \emph{academy\_3\_vs\_1\_with\_keeper}. \textbf{Middle:} \emph{academy\_pass\_and\_shoot\_with\_keeper}. \textbf{Right:} \emph{academy\_corner}.}
    \label{fig:court}
\end{figure*}

\section{Implementation Details}

\subsection{Viewpoint Invariance for Moving Viewpoint}

\begin{figure}[htbp]
    \centering
    \subfigure{
    \includegraphics[width=1.8 in]{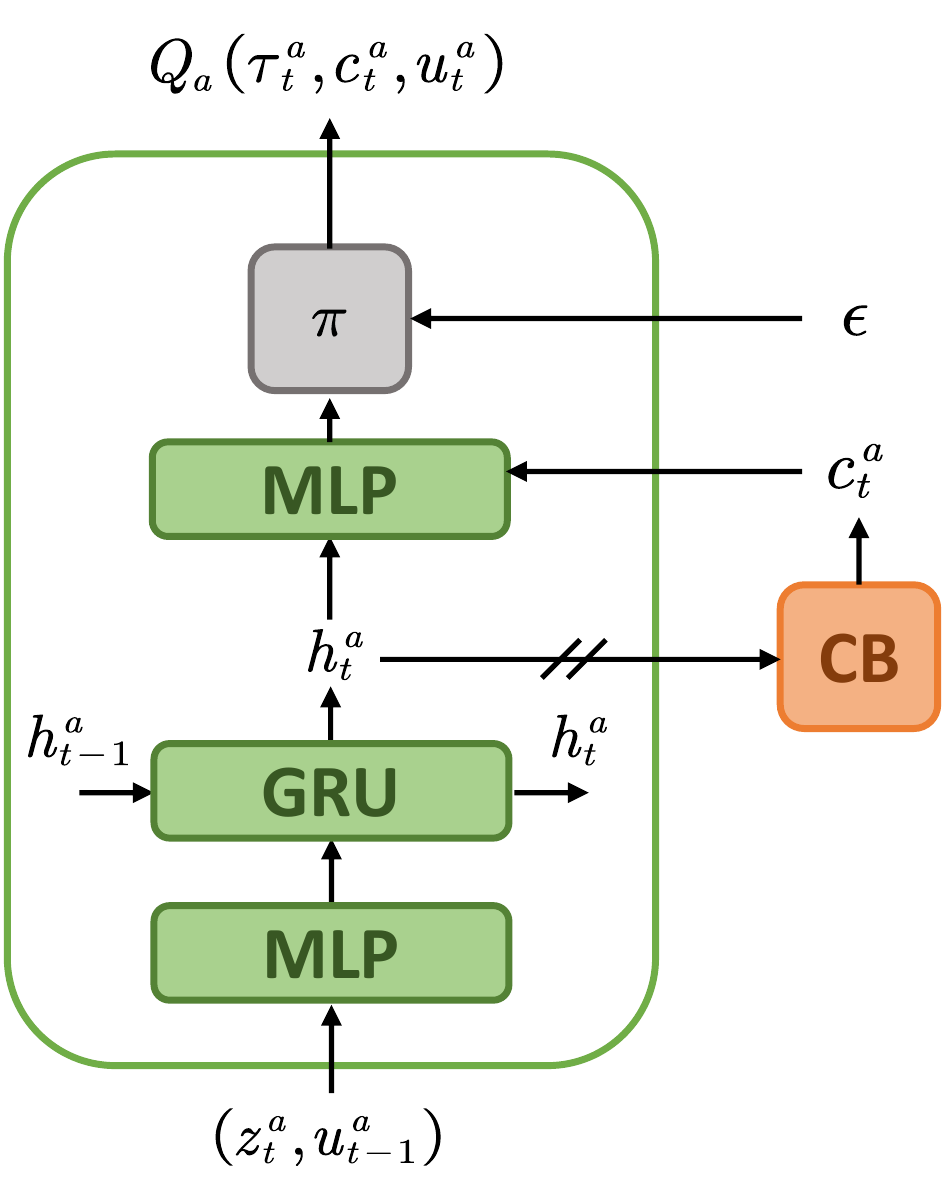}
    }
    \hspace{0.2 in}
    \subfigure{
    \includegraphics[width=4.4 in]{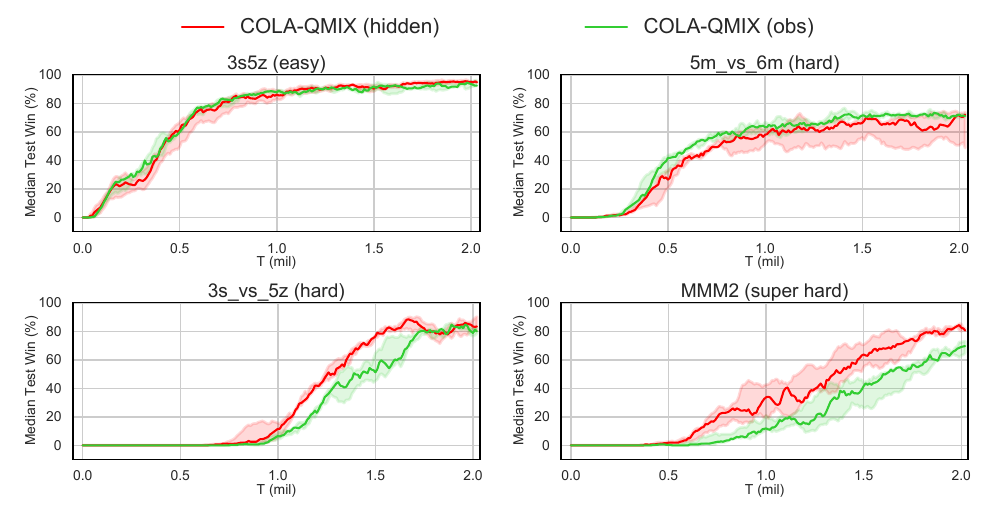}
    }
\caption{\textbf{Left:} Alternative architecture with $h_t^a$ as the input. \textbf{Right:} Performance comparison in SMAC.}
\label{fig:obs}
\end{figure}

Unlike traditional multi-view problems, agents in most multi-agent systems can move around. Moving agents indicate that the viewpoints are not fixed. This might make it more challenging to train the consensus builder, especially in partially observable environments. To mitigate this problem, we replace the original local observations with historical trajectories as the input of the consensus builder. For the value decomposition method, we regard the hidden output $h_t^a$ of the recurrent neural network in the agent network as the integration of all historical information. It should be noted that we apply a stop-gradient operator on $h_t^a$ for the input of the consensus builder. The structure of the modified agent network in the value decomposition variant with consensus learning is shown in Figure~\ref{fig:obs}. 

We compare the performance of the consensus builder when the input is $h_t^a$ or raw local observations. Figure~\ref{fig:obs} presents that COLA-QMIX with historical information as input performs slightly better. For the multi-agent actor-critic method, we keep local observations as input. And as shown in Figure~\ref{fig:mpe_result}, COLA-MADDPG can still improve the original algorithm.

\subsection{Algorithmic Description}

The algorithms for training multi-agent reinforcement learning methods with consensus learning are summarized in Algorithm~\ref{alg:COLA-QMIX} and Algorithm~\ref{alg:COLA-MADDPG}. The code for COLA can be found in the supplementary material.

\begin{algorithm}[htbp]
\caption{Training Procedure for COLA-QMIX}
\label{alg:COLA-QMIX}
\textbf{Hyperparameters}: $K$, $\gamma$, $\epsilon$\\
Initialize the parameters of the agent network and the mixing network\\
Initialize the parameters of the consensus builder
\begin{algorithmic}[1] 
\FOR{each episode}
\STATE Obtain the global state $s_1$ and the local observations $\boldsymbol{z}_1=\{z^1_1, z^2_1,\dots, z^n_1\}$
\FOR{$t \leftarrow 1$ to $T-1$}
\FOR{$a \leftarrow 1$ to $n$}
\STATE Calculate the consensus $c_t^a$ according Eq.~\eqref{eq:consensus}
\STATE Select action $u^a_t$ according to $\epsilon$-greedy policy w.r.t $Q_a(\tau^a_t, c^a_t, \cdot)$
\ENDFOR
\STATE Take the joint action $\boldsymbol{u}_t=\{u_t^0, u_t^1,\dots,u_t^n\}$
\STATE Obtain the global reward $r_{t+1}$, the next local observations $\boldsymbol{z}_{t+1}$, and the next state $s_{t+1}$
\ENDFOR
\STATE Store the episode in $\mathcal{D}$
\STATE Sample a batch of episodes $\sim$ Uniform($\mathcal{D}$)
\STATE Update the parameters of the consensus builder according Eq.~\eqref{eq:cb_loss}
\STATE Update the parameters of the agent network and the mixing network according Eq.~\eqref{eq:ctd_loss}
\STATE Replace target parameters every $M$ episodes
\ENDFOR
\end{algorithmic}
\end{algorithm}
\begin{algorithm}[htbp]
\caption{Training Procedure for COLA-MADDPG}
\label{alg:COLA-MADDPG}
\textbf{Hyperparameters}: $K$, $\gamma$\\
Initialize the parameters of the actor and the critic\\
Initialize a random process $\mathcal{N}$ for action exploration\\
Initialize the parameters of the consensus builder
\begin{algorithmic}[1] 
\FOR{each episode}
\STATE Obtain the local observations $\boldsymbol{z}_1=\{z^1_1, z^2_1,\dots, z^n_1\}$
\FOR{$t \leftarrow 1$ to $T-1$}
\FOR{$a \leftarrow 1$ to $n$}
\STATE Calculate the consensus $c_t^a$ according Eq.~\eqref{eq:consensus}
\STATE Execute action $u^a_t=\mu_{a}(z^a_t, c^a_t) + \mathcal{N}_t$
\ENDFOR
\STATE Take the joint action $\boldsymbol{u}_t=\{u_t^1, u_t^2,\dots,u_t^n\}$
\STATE Obtain the global reward $r_{t+1}$ and the next local observations $\boldsymbol{z}_{t+1}$
\STATE Store $(\boldsymbol{z}_t, \boldsymbol{u}_t, r_{t+1}, \boldsymbol{z}_{t+1})$ in replay buffer $\mathcal{D}$
\FOR{$a \leftarrow 1$ to $n$}
\STATE Sample a minibatch of $N$ samples $(\boldsymbol{z}, \boldsymbol{u}, r, \boldsymbol{z}^\prime)$ $\sim$ Uniform($\mathcal{D}$)
\STATE Calculate consensuses $c^a$ and ${c^\prime}^a$ according to Eq.~\eqref{eq:consensus}
\STATE Calculate $y=r+\gamma Q_a^{\boldsymbol{\mu}^\prime}\left.\left(\boldsymbol{z}^\prime, {c^\prime}^a, {u^\prime}^1, {u^\prime}^2, \dots, {u^\prime}^n\right)\right|_{{u^\prime}^i={\boldsymbol{\mu}^\prime}^i({z^\prime}^i, {c^\prime}^i), i \in \{1, 2, \dots, n\} }$
\STATE Update critic by minimizing the loss $\mathcal{L}=\frac{1}{N}\sum\left(y - Q_a^{\boldsymbol{\mu}}\left(\boldsymbol{z}, c^a, \boldsymbol{u}\right)\right)^2$
\STATE Update actor according to Eq.~\eqref{eq:policy_gradient}
\ENDFOR
\STATE Update target parameters by polyak averaging
\ENDFOR
\ENDFOR
\end{algorithmic}
\end{algorithm}

\subsection{Hyperparameters}

The hyperparameters we employ for various algorithms are listed in Table~\ref{table:hyperparameters}. The hyperparameters of variants with consensus learning are consistent with the original algorithms. We make sure that the hyperparameters of other baselines are the same as the ones in their official code.

\begin{table}[h]
\centering
\begin{tabular}{lll}
\hline
\textbf{Algorithm}                 & \textbf{Description}                                        & \textbf{Value} \\ \hline
\multirow{6}{*}{QMIX}              & Type of optimizer                                           & RMSProp        \\
                                   & Learning rate                                               & 0.0005         \\
                                   & How many episodes to update target networks                 & 200            \\
                                   & Batch size                                                  & 32             \\
                                   & Capacity of replay buffer (in episodes)                     & 5000           \\
                                   & Discount factor $\gamma$                                    & 0.99           \\ \hline
\multirow{5}{*}{MADDPG}            & Type of optimizer                                           & Adam           \\
                                   & Learning rate                                               & 0.01           \\
                                   & Batch size                                                  & 1024           \\
                                   & Capacity of replay buffer                                   & 1000000        \\
                                   & Discount factor $\gamma$                                    & 0.95           \\
                                   & How many steps to update networks                           & 100            \\ \hline
\multirow{7}{*}{Consensus Builder} & Number of categories of consensus $K$ for MPE               & 4              \\
                                   & Number of categories of consensus $K$ for SMAC              & 4 or 32        \\
                                   & Number of categories of consensus $K$ for GRF               & 32             \\
                                   & Temperature parameter of teacher networks $\tau_{T}$        & 0.04           \\
                                   & Temperature parameter of student networks $\tau_{S}$        & 0.1            \\
                                   & Smoothing parameter in the update rule for the center       & 0.9            \\
                                   & Smoothing parameter in the update rule for teacher networks & 0.996          \\ \hline
\end{tabular}
\caption{Hyperparameter settings.}
\label{table:hyperparameters}
\end{table}

\section{Additional Experiments}

\begin{figure*}[h]
    \centering
    \subfigure{
    \includegraphics[width=0.45\linewidth]{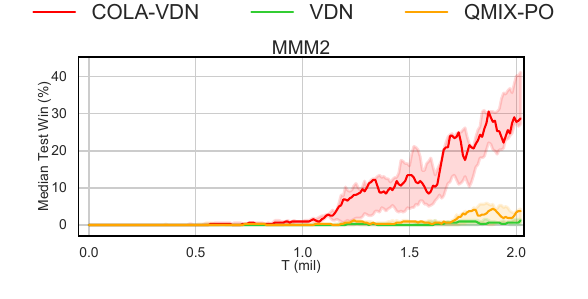}
    }
    \hspace{0.2 in}
    \subfigure{
    \includegraphics[width=0.45\linewidth]{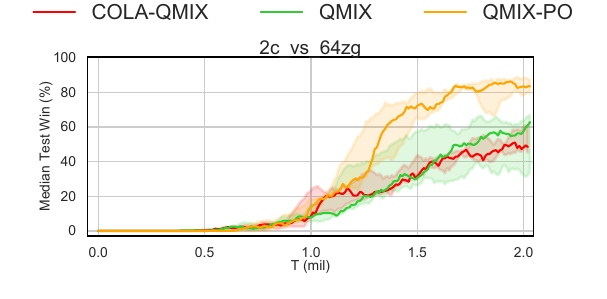}
    }
    \caption{\textbf{Left}: Learning curves of our COLA-VDN and other algorithms on \emph{MMM2}. Note that none of the algorithms can get the true global state. \textbf{Right}: Learning curves of several algorithms in \emph{2c\_vs\_64zg} scenarios where global state information is not important.}
    \label{fig:addition_experiment}
\end{figure*}

We apply the COLA framework to VDN and evaluate its performance in the super hard scenarios \emph{MMM2} on SMAC. Due to the fact that neither COLA-VDN nor the original algorithm VDN use the global state, the consensus signal generated by the consensus builder is crucial for promoting cooperation between agents. Under the assumption that the perfect state information cannot be obtained, we concatenate the local observations of our own agents and input them as the state to QMIX. And we call this variant QMIX-PO. As can be seen from Figure~\ref{fig:addition_experiment}, neither VDN nor QMIX-PO can successfully solve the task when the global state is inaccessible. Without using the global state, COLA-VDN is significantly better than other algorithms by inferring the consensus signal corresponding to the state.

On the contrary, in some scenarios where the global state is not important, the improvement brought by the COLA framework is not obvious. In \emph{2c\_vs\_64zg} map, due to the large number of enemy units, agents should only focus on the enemies in view. QMIX-PO in Figure~\ref{fig:addition_experiment}, which cannot obtain the global state, achieves a state-of-the-art performance in \emph{2c\_vs\_64zg} scenario. Therefore COLA does not perform well in a few scenarios where the agents do not need to reach consensus on the global state.

\end{document}